# A Boundary-Spheropolygon Element Method for Modelling Sub-particle Stress and Particle Breakage

YUPENG JIANG[1], HANS J. HERRMANN[2], FERNANDO ALONSO-MARROQUIN*[1],

1. *School of Civil Engineering, The University of Sydney, Australia*
2. *Computational Physics for Engineering Materials, ETH Zurich, Switzerland*

Abstract: We present a boundary-spheropolygon element method that combines the boundary element method (BEM) and the spheropolygon-based discrete element method (SDEM). The interaction between particles is simulated via the SDEM, and the sub-particle stress is calculated by BEM. The framework of BSEM is presented. Then, the accuracy and efficiency of the method are analysed by comparison with both analytical solutions and a well-established finite element method. The results demonstrate that BSEM can efficiently provide instant sub-particle stress for particles with an optimized compromise between computational time and accuracy, thus overcoming most of the disadvantages of existing numerical methods.

## 1. Introduction

Particle breakage is ubiquitous for granular materials. Breakage is caused by many reasons, such as strong compressive loading, large shear displacement, extreme temperature, dehydration or chemical effects. Based on laboratory studies, it is concluded that breakage causes changes in particle size distribution [1-4] and particle shape [5,6], which in turn change the state variables, such as density and compressibility, that govern the mechanical behaviour of the bulk granular material. These studies have provided a framework for studying particle breakage. However, due to the limitations in experimental observations, laboratory studies are not able to offer more detailed information about the breakage process.

The discrete element method (DEM), alternatively, provides an insightful way to numerically study the breaking process. It allows the particle breakage to be examined at the grain scale, and the breaking process to be captured at the smallest time scale. The strong and weak force chains can be visualized with DEM, and the breakage of each particle can be traced and analysed. For these reasons, DEM has been widely applied to study particle breakage. Existing methods have been summarized into two paradigms: agglomerates and replaceable particles [7]. Agglomerates [8-12] involve representing individual particles by groups of smaller sub-particles that are bonded together through pre-existing forces. The bonds will be broken or, from the computational perspective, erased once their force reaches a certain threshold. The replaceable particles, alternately, simulate each grain as a single entity and replace them with smaller fragments when the characteristic stress within the original entity exceeds the particle strength.

The agglomerates can dynamically simulate fracturing processes of individual particles, but the numerical implementation is highly inefficient. This is because simulations require a large number of sub-particles to approach the geometrical shape of particles and obtain reliable results. The fact that the smallest particle size is limited to the size of non-breakage elements compromises its ability to generate a correct particle size distribution (PSD). Unrealistic voids are also inevitably introduced if circular particles are used.

The replaceable particle method is more desirable over agglomerates [7]. The earliest study was conducted by Åström and Herrmann [13] using contact force criteria. Tsoungui et al. [14] proposed a model for crushing circular grains with average stress tensor criteria. They developed a decomposition method to define the failure of an individual grain subjected to an arbitrary set of contact forces. Cantor et al. [15] further modified the method for irregular particles' fragmentation. Eliáš [16] and Gladkyy & Kuna [17] developed similar stress-based crushable models with three-dimensional polyhedral particles; fragmentation of grains is included by splitting the particles into certain patterns of smaller polyhedrons.

High computational efficiency is the greatest advantage of replaceable particles. Meanwhile, such an approach also avoids the inherent problems in determining the current void ratio for an aggregate of disks, which are porous and have internal voids [7]. However, a correct local stress distribution and the conservation of mass cannot be guaranteed if circular particles are adopted. The replacement strategy lacks solid physical meaning. Meanwhile, the reliability and efficiency of the simulation are largely compromised if angular particles are adopted, for which the relation of contact forces becomes much more complicated and replacement strategies are poorly studied. It can be observed from the results of these studies [16,17] that the ultimate PSD is quite unsatisfying. This is because the average stress method, which is widely used for circular cases, becomes less valid when shape effects are not negligible. Both breakage initiation and shapes of new fragments are extremely difficult to determine with an averaged stress tensor.

A variety of continuous-discontinuous methods have been proposed to fulfil the breakage requirements [18,19]. The basic idea of these hybrid methods is to combine different types of continuous techniques with DEM. The grain-to-grain interaction is simulated via DEM, and the stress state of an individual particle is calculated by continuous methods using contact force information borrowed from DEM. However, existing methods are still not able to provide results with both high accuracy and reasonable efficiency. For example, the finite-discrete element method [19] can accurately calculate the stress state and breakage surface, but it is computationally very expensive. High-quality results require a high number of elements in the mesh on an individual particle. FEM also bears the same disadvantage of agglomerates that is inherited due to the unbreakable finite elements. Luo et al. [18] proposed a combined scaled boundary finite-discrete element method, which is quite efficient; however, it is unable to provide an accurate stress field in each individual particle, and the calculation for the breakage interface is also oversimplified.

In this manuscript, we present a new method, the boundary-spheropolygon element method (BSEM), which combines the advantages of the continuum and discrete methods. It currently consists of a one-way coupling between SDEM and BEM. The sub-particle stress is calculated through the boundary element method, and the interaction between particles is computed with the spheropolygon-based discrete element method. BSEM only requires the mesh at the boundary of particles and thus simplifies the mesh generation. It uses the fundamental solutions for stress problems. The numerical errors are only generated at the boundary, while the displacement and stress field inside the particles are infinitely differentiable. BSEM has a relatively higher efficiency for calculations with a lower number of degrees of freedom, and it is computationally less expensive than FEM for the same accuracy. The irregularity of particles and the contact information are effectively captured through the spheropolygon element method. This new method is intended to be the first step towards the development of a new mechanistic approach for grain breakage based on an evaluation of the sub-particle stress.

The paper is organized as follows. The computational methodology of BSEM is presented in Section 2. The validation and performance of the method are discussed in Section 3 with both analytical solutions and FEM results of Brazilian tests. The single-particle tests for BSEM are conducted through a series of parametric studies on the effect of aspect ratio, coordination number, and heterogeneity of the sub-particle stress. In Section 4, the BSEM is applied to uniaxial compression tests in granular models to investigate the relation between the particle circularity and the breakage resistance. A general discussion and conclusions are presented in Section 5.

## 2. Boundary-spheropolygon element method (BSEM)
2.1 Boundary element method (BEM)

The boundary element method (BEM) is used as the continuous method to calculate the sub-particle stress. This method subdivides the boundary of the particle under consideration into a series of elements [20]. The solution is sought as a linear combination of fundamental solutions, and their coefficients are solved by satisfying the boundary conditions through boundary integral equations. Existing literature has proved that BEM can effectively solve problems for elasticity. However, it still has not been combined with DEM in the way we present here.

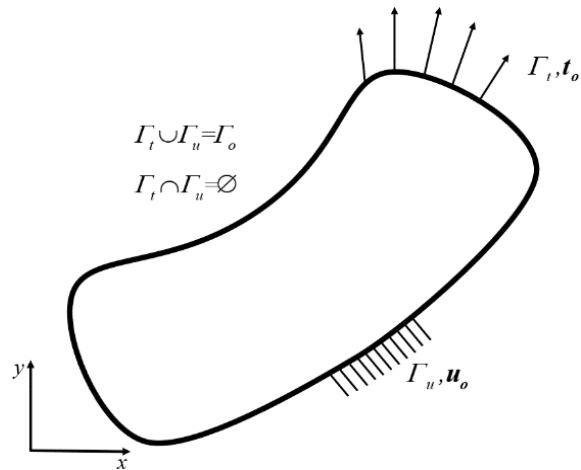

Figure 1 Definition of boundary conditions in BEM. $u_0$ and $t_0$ are the boundary conditions for displacement and traction, respectively. $\Gamma_u$ and $\Gamma_t$ denote the surface area for the displacement/traction condition that is known, and they cover the entire boundary of the domain $\Omega$.

In two-dimensional granular flow, the quasi-static sub-particle stress can be regarded as a plane elastostatic problem. Let us consider a homogenous area as shown in Figure 1. The boundary conditions can be written as:

$$u(x)=u_0, \quad x \in \Gamma_u \quad \text{(Dirichlet boundary conditions)}, \tag{1}$$

$$t(x)=t_0, \quad x \in \Gamma_t \quad \text{(Neumann boundary conditions)}, \tag{2}$$

where $u$ ($u_1$, $u_2$) and $t$ ($t_1$, $t_2$) denote the boundary displacements and boundary tractions, respectively. Somigliana's identity, which represents the relationship between a domain point and the boundary conditions, is used to move the domain point to the boundary. The boundary integral equation is expressed as [20,21-23]:

$$c_{ij}u_j(p)=\int_\Gamma u^*_{ij}(p,q)t_j(q)\mathrm{d}\Gamma - \int_\Gamma t^*_{ij}(p,q)u_j(q)\mathrm{d}\Gamma + \int_\Omega b_j u^*_{ij}(p,q)\mathrm{d}\Omega, \tag{3}$$

where $u^*_{ij}(p,q)$ and $t^*_{ij}(p,q)$ are the Kelvin fundamental solutions. They denote the $i$th component of displacement and traction at point $q$ induced by the unit load at point $p$ in the $j$th direction. The vector $b$ is the body force, and $c_{ij}$ represent the boundary coefficients that can be written as $c_{ij}=\delta_{ij}/2$ with the Kronecker delta $\delta_{ij}$. All of the subscripts ($i$, $j$, $k$, $l$=1,2) follow the Einstein notation rule. The Kelvin fundamental solutions for the displacements and boundary tractions for an isotropic body are written as [24,25]:

$$u^*_{ij}(p,q) = \frac{1}{8\pi G(1-\nu)}[(3-4\nu)\ln(\frac{1}{r})\delta_{ij} + r_{,i}r_{,j}], \tag{4}$$

$$t^*_{ij}(p,q)= \frac{1}{4\pi(1-\nu)r}\{r_{,n}[(1-2\nu)\delta_{ij} + 2r_{,i}r_{,j}] - (1-2\nu)(r_{,i}n_j - r_{,j}n_i)\}, \tag{5}$$

where $G$ is the shear modulus and $\nu$ denotes Poisson's ratio. The scalar $r$ is the Euclidean distance from $p$ to $q$, and $r_{,i} = (p_i-q_i)/r$ denote the derivatives with respect to $p_i$. The vector component $n_i$ is the direction cosine of the unit outward normal vector at $q$. The quantity $r_{,n}=r_{,i}n_i$ represents the derivative of $r$ in the direction of the outward normal at the point $p$.

2.1.1 Boundary discretization and integration

Based on the boundary integral equation, the boundary can be discretized with a finite number of elements, as shown in Figure 2. In the absence of a body force term, the discretised equation for a boundary element can be written as:

$$c_{ij}u_j(p)=\sum_{l=1}^{N}[\int_{\Gamma_l} u^*_{ij}(p,q)t_j(q)\mathrm{d}\Gamma - \int_{\Gamma_l} t^*_{ij}(p,q)u_j(q)\mathrm{d}\Gamma]. \tag{6}$$

Then, all of the boundary elements form a set of linear equations:

$$c_m u_m = \sum_{l=1}^{N}\{\int_{\Gamma_l}(u^*_{lm}t_l - t^*_{lm}u_l)\mathrm{d}\Gamma\}, \tag{7}$$

where $l$ and $m$ are the references of boundary elements, and $N$ is the total number of elements ($l$,

$m=1……N$). The constant element is used to achieve optimal efficiency, which means $u=u_i$ and $t=t_i$ ($i=1,2$) are constant values within each element.

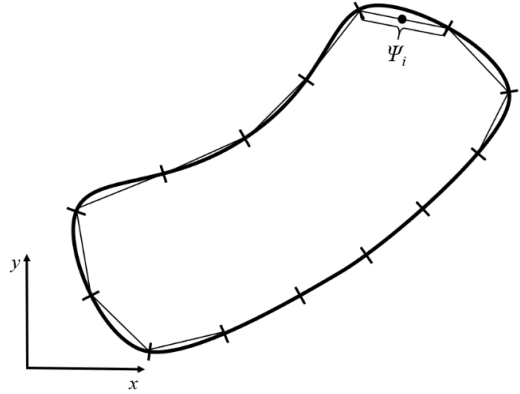

Figure 2 The boundary of the domain is discretised into a series of piecewise lines, and each line is regarded as a boundary element $\Psi_i$.

The matrices of the fundamental solutions are:

$$\boldsymbol{t}^* = \begin{bmatrix} t^*_{11} & t^*_{12} \\ t^*_{21} & t^*_{22} \end{bmatrix}, \quad \boldsymbol{u}^* = \begin{bmatrix} u^*_{11} & u^*_{12} \\ u^*_{21} & u^*_{22} \end{bmatrix}. \tag{8}$$

$$\boldsymbol{H}_{lm} = (\boldsymbol{c}_m + \int_{\Gamma_l} \boldsymbol{t}^*_{lm} \mathrm{d}\Gamma), \tag{9}$$

$$\boldsymbol{G}_{lm} = \int_{\Gamma_l} \boldsymbol{u}^*_{lm} \mathrm{d}\Gamma, \tag{10}$$

Therefore, the matrix for the unknown boundary elements is:

$$\sum_{l=1}^{N} \boldsymbol{H}_{lm} \boldsymbol{u}_l = \sum_{l=1}^{N} \boldsymbol{G}_{lm} \boldsymbol{t}_l, \tag{11}$$

$$\boldsymbol{AU} = \boldsymbol{BT}, \tag{12}$$

where $\boldsymbol{U}^T = [\boldsymbol{u}^T_1, \boldsymbol{u}^T_2, \boldsymbol{u}^T_3, ……, \boldsymbol{u}^T_N]$ and $\boldsymbol{T}^T = [\boldsymbol{t}^T_1, \boldsymbol{t}^T_2, \boldsymbol{t}^T_3, ……, \boldsymbol{t}^T_N]$ are the nodal displacements and nodal tractions; the coefficient matrices $\boldsymbol{A}_{2N \times 2N}$ and $\boldsymbol{B}_{2N \times 2N}$ are constructed as block matrices of the 2×2 matrices $\boldsymbol{H}$ and $\boldsymbol{G}$ given by Eqs. (9) and (10). After inverting $A$ and obtaining all of the boundary displacements, the stress state at any position of the particle can be calculated as:

$$\sigma_{ij} = \int_{\Gamma} D_{kij} t_k \mathrm{d}\Gamma - \int_{\Gamma} S_{kij} u_k \mathrm{d}\Gamma, \tag{13}$$

where the tensor components of $D_{kij}$ and $S_{kij}$ are given as [20,25]:

$$D_{kij} = \frac{1}{r}\{(1-2\nu)[\delta_{ki}r_{,j} + \delta_{kj}r_{,i} - \delta_{ij}r_{,k}] + 2r_{,i}r_{,j}r_{,k}\}\frac{1}{4\pi(1-\nu)}, \tag{14}$$

$$S_{kij} = \frac{2}{r^2}\left\{\begin{array}{l} 2\dfrac{\partial r}{\partial n}[(1-2\nu)\delta_{ij}r_{,k} + \nu(\delta_{ik}r_{,j} + \delta_{jk}r_{,i}) - 4r_{,i}r_{,j}r_{,k}] \\ +2\nu(n_i r_{,j}r_{,k} + n_j r_{,i}r_{,k}) + (1-2\nu)(2n_k r_{,i}r_{,j} + n_j\delta_{ik} + n_i\delta_{ik}) \\ -(1-4\nu)n_k\delta_{ij} \end{array}\right\}\frac{1}{4\pi(1-\nu)}. \tag{15}$$

A four-point Gaussian-Legendre quadrature is adopted as the numerical integration method in both Eqs. (9) and (13).

2.1.2 Regularization

The method of *regularization* is a highly effective way to eliminate the rigid-body mode in a coefficient matrix (i.e., matrix $A$ in Eq. (12)) of elasticity. The major issue in calculating sub-particle stress is that, for most of the particles, all boundary tractions are given, and all boundary displacements are unknown. Thus, the solution is not unique under pure Neumann boundary conditions. The free rigid-body mode (RBM) is inevitably provoked, and it makes the accuracy of the solution highly compromised or even incorrect. The usual measure for eliminating the RBM is to restrain the body, which requires the introduction of extra conditions. However, restraining is not an ideal method for the particles in DEM simulations where the particles are usually free to exhibit rigid-body movement and rotation. The alternative method of *regularization* is used to eliminate the effect of RBM and obtain the stress field. This method does not require the definition of restraining conditions or any extra procedure of the mesh. The stress components at each time step can be calculated instantly with the DEM simulation.

The regularization method for the non-symmetric matrix was proposed by Lutz [28]. It involves first computing the singular stiffness matrix and then suitably modifying it using linear algebra. Considering the non-symmetric square matrix in $AU=BT$ (Eq. (9)), the $B$ matrix must be non-singular since the traction $T$ is uniquely defined by the displacement $U$; the null space $N(A)$ of $A$, however, is spanned by the rigid-body modes since pure traction conditions are not able to define a unique solution. The basis of $N(A)$ for the two-dimensional case is written as:

$$\gamma_1^T = [1, 0, 1, 0, 1, 0, ......., 1, 0],$$

$$\gamma_2^T = [0, 1, 0, 1, 0, 1, ......., 0, 1],$$

$$\rho^T = [-y_1, x_1, -y_2, x_2, ........, -y_N, x_N], \tag{23}$$

where $\gamma_1$, $\gamma_2$ and $\rho$ denote two translation vectors and a rotation vector, respectively. The solution for
$$Ax=b, \tag{24}$$
can be written as
$$x=z+\eta y, \tag{25}$$
where $z$ is orthogonal to the null space $\zeta_i=\gamma_i$ ($i=1,2$) and $\zeta_i=\rho_i$ ($i=3$), which consist of the matrix $\eta_{2N\times 3}$; $y$ is the coefficient matrix. Now, define a new matrix $C$ as
$$C=A+J\eta^T, \tag{26}$$
with $J_{2N\times 3}$ and the rank of $J$ equal to three. Obviously,
$$Cz=(A+J\eta^T)z = Az+0 = b, \tag{27}$$
where the vector $z$ is the solution without the rigid body mode. The form of $J$ for a nonsymmetric matrix is written as:

$$\boldsymbol{J}=[\boldsymbol{d}_1,\boldsymbol{d}_2,\boldsymbol{d}_3], \qquad \boldsymbol{d}_i=\boldsymbol{B}\boldsymbol{\xi}_i. \tag{28}$$

The rigid body mode could also be removed from the solution after solving the matrix $A$ with a non-zero nullity.

2.2 Spheropolygon discrete element method (SDEM)

The spheropolygon discrete element method (SDEM) is selected as the DEM part of the combined algorithm due to its unique advantages [26, 27]. The geometrical irregularity of a particle is properly represented as a Minkowski sum of a polygon with a disk, and the multiple contacts between irregular particles are calculated based on distances between vertices and edges. These features make BSEM computationally efficient. The contact's information can be easily borrowed by the traction vector as boundary conditions in the continuous part of BSEM. Most importantly, point contacts are more realistic loading conditions for particles prior to breakage in granular materials.

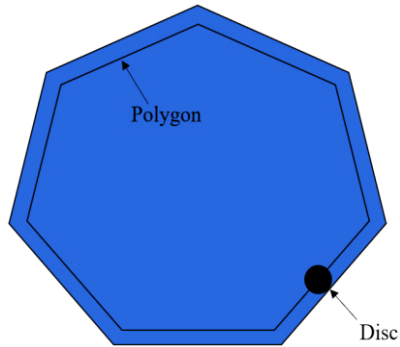

Figure 3 Spheropolygon element obtained by sweeping a disk around a polygon.

A spheropolygon is the Minkowski sum of a polygon with a certain number of vertices representing the irregular shape and a disk with radius $r$, which defines the elastic area for the calculation of contact forces. Formally, the Minkowski sum of two sets of points $P$ and $Q$ of a vector space is given by:
$$P+Q = \{\boldsymbol{x}+\boldsymbol{y}|\ \boldsymbol{x}\in P, \boldsymbol{y}\in Q\} \tag{16}$$
This operation is geometrically equivalent to the sweeping of a disc around the profile of the polygon without changing its relative orientation [26]. For example, the irregular shape of the polygon in Figure 3 can be accurately approximated by a spheropolygon element with a few boundary lines and a disc sweeping its profile. The polygon defines the element shape, and the disc is used for contact calculation. This is a more effective approach than using a cluster of small particles to approach this shape, especially when the particle shape is more complicated. A variety of complex shapes can be generated using spheropolygons, which can properly represent the irregular shape of particles for a simulation of breakage with optimal efficiency.

The interaction force of the spheropolygon is defined by vertex-edge forces. For example, let us consider two spheropolygons $SP_i$ and $SP_j$ with their polygons $P_i$ and $P_j$ and the radii of the disks $ra_i$ and $ra_j$. Each polygon is defined by its own set of vertices $\{V\}$ and edges $\{E\}$. The overlapping length between each vertex-edge pair $(V, E)$ is written as:

$$\delta(V,E) = \langle ra_i + ra_j - d(V,E) \rangle, \tag{17}$$

where $d(V, E)$ is the Euclidean distance between the vertex $V$ and the edge $E$. Therefore, the force applied on particle $i$ by particle $j$ is expressed as:

$$\boldsymbol{F}_{ij} = -\boldsymbol{F}_{ji} = \sum_{V_i E_j} \boldsymbol{F}(V_i, E_j) + \sum_{V_j E_i} \boldsymbol{F}(V_j, E_i), \tag{18}$$

and the torque on particle $i$ is:

$$\boldsymbol{\tau}_{ij} = \sum_{V_i E_j} (\boldsymbol{p}(V_i, E_j) - \boldsymbol{s}_i) \times \boldsymbol{F}(V_i, E_j) + \sum_{V_j E_i} (\boldsymbol{p}(V_j, E_i) - \boldsymbol{s}_i) \times \boldsymbol{F}(V_j, E_i), \tag{19}$$

where $s_i$ is the mass centre of particle $i$ and $\boldsymbol{p}$ is the point of applied force, which is defined as the middle point of the overlap area between the vertex and the edge:

$$\boldsymbol{p}(V,E) = \boldsymbol{X} + (ra_i - \frac{1}{2}\delta(V,E))\frac{\boldsymbol{X}-\boldsymbol{Y}}{\|\boldsymbol{Y}-\boldsymbol{X}\|}, \tag{20}$$

where $\boldsymbol{X}$ is the position of the vertex V, and $\boldsymbol{Y}$ is its closest point on the edge E. The movement of the centre of mass $r_i$ and the orientation $\varphi_i$ of the particle are governed by the equations of motion:

$$m_i \ddot{\boldsymbol{s}} = \sum_j \boldsymbol{F}_{ij} + m_i g \boldsymbol{n}, \qquad I_i \ddot{\boldsymbol{\varphi}} = \sum_j \boldsymbol{\tau}_{ij}, \tag{21}$$

where $m_i$ and $I_i$ are the mass and moment of inertia of the particle; $g$ denotes the gravitational acceleration; and $\boldsymbol{n}$ is the unit vector along the direction of gravity. The linear elastic model is used throughout this paper. The force $\boldsymbol{F}$ can be written as:

$$\boldsymbol{F} = k_n \delta_n \boldsymbol{N} + k_t \delta_t \boldsymbol{T}, \tag{22}$$

where $k_n$ and $k_t$ are the normal and tangential stiffness, respectively; $\boldsymbol{N}$ and $\boldsymbol{T}$ are the normal and tangential unit vector; $\delta_n$ denotes the length of the overlap; and $\delta_t$ is the elastic displacement that accounts for frictional forces [26].

2.3 Coupling between BEM and SDEM

The coupling process between the BEM and SDEM is mainly conducted by importing contact forces into the traction vector (Eq. (12)). Currently, the BSEM is a one-way coupling method that only calculates the sub-particle stress but does not consider the breakage of the particle. The fully coupled BSEM should further replace the broken particles [7] at each time step based on their stress state. In this subsection, we introduce the details of coupling procedures. General considerations for accuracy and computational efficiency are also discussed, which include methods for avoiding the near-boundary singularity and necessary conditions to guarantee the quasi-static state of a particle before conducting the stress calculation.

2.3.1 Avoid the near-boundary singularity

The *near-boundary singularity* refers to the divergences that occur when an integral point of BEM approaches the boundary. The error of numerical integrands becomes quite large over the length of the element [29]. This problem is caused by fundamental solutions, which involve distance-dependent terms such as $\ln(1/r)$, $1/r$, and $1/r^2$. There are two methods to avoid or minimize such a problem. The first approach is to use a parameterized Gaussian quadrature method [29, 30] when a region point is approaching the boundary. The other method is to properly ignore the near-boundary area through the particle dilation process of the SDEM. It is concluded from existing literature [10,11,31] that the breakage is mainly caused by tensile failures inside the particle. Therefore, the exact values of boundary displacement may not be crucial for particle fragmentation, and the analysis of Cauchy principal values for the particles is deemed unnecessary.

SDEM requires the polygon, which is the rigid part of the particle that defines its shape, to be dilated by the half-diameter size of its disc so that an elastic contact layer can be created with its disc. Alarcon et al. [20] indicated that the near-field singularity of the integrand vanishes at approximately the half-distance of an element. This strategy is incorporated as follows through particle dilation in BSEM to reduce the effect of the near-field singularity: (i) the length of the element and the radius of the spheropolygon disc are equal, (ii) the boundary for integration is dilated from the rigid one (polygon) to the elastic boundary (polygon + radius), and (iii) only the stress field within the polygon is trusted and used for breakage calculation. Figure 6 shows how the polygon is dilated for the calculation of sub-particle stress; the boundary discretization is conducted at the dilated boundary.

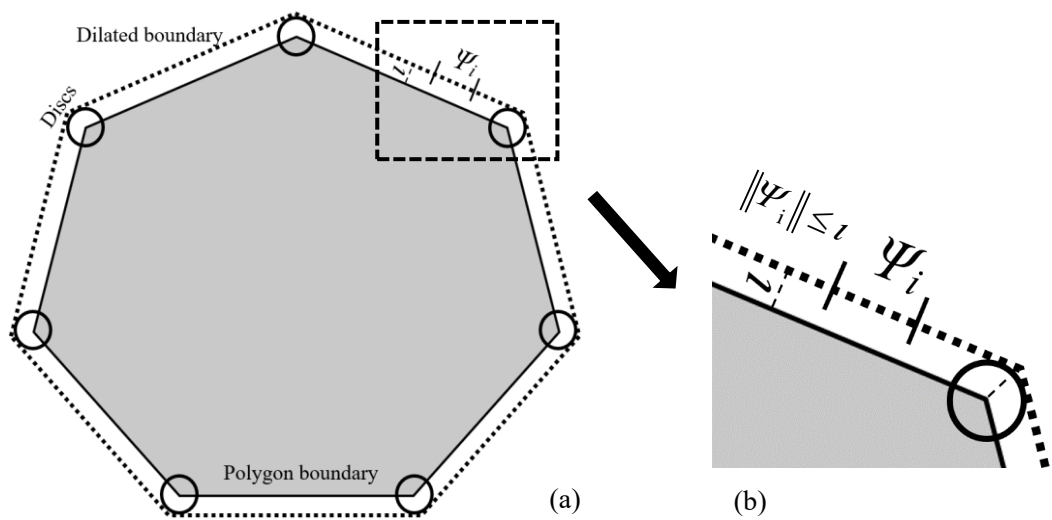

Figure 6 The particle dilation method is used to avoid the near-boundary singularity. (a) The Minkowski sum of the polytonal particle and disk of radius $\iota$; (b) enlarged square area in (a) at the boundary. If the length of the element $\Psi_i$ is lower than or equal to $\iota$, the near-boundary singularity inside the polygon can be ignored.

By using particle dilation in BSEM, the accuracy of the stress state can be guaranteed. It ignores the near-singular area through the dilation without losing too much stress information inside the polygon. It is computationally more efficient than the parameterized Gaussian quadrature. Although the actual attrition and erosion near the particle boundary may not be accurately simulated, the fragmentation of particles can be properly captured through the sub-particle stress fields inside the polygon.

2.3.2 Coupling through contact forces

In BSEM, the interaction between spheropolygons is calculated from vertex-edge contact forces. Each contact is defined by a vector representing the contact force and a vector representing the point of application of the contact force. This information needs to be passed to the BEM for the calculation of the sub-particle stress as a boundary problem.

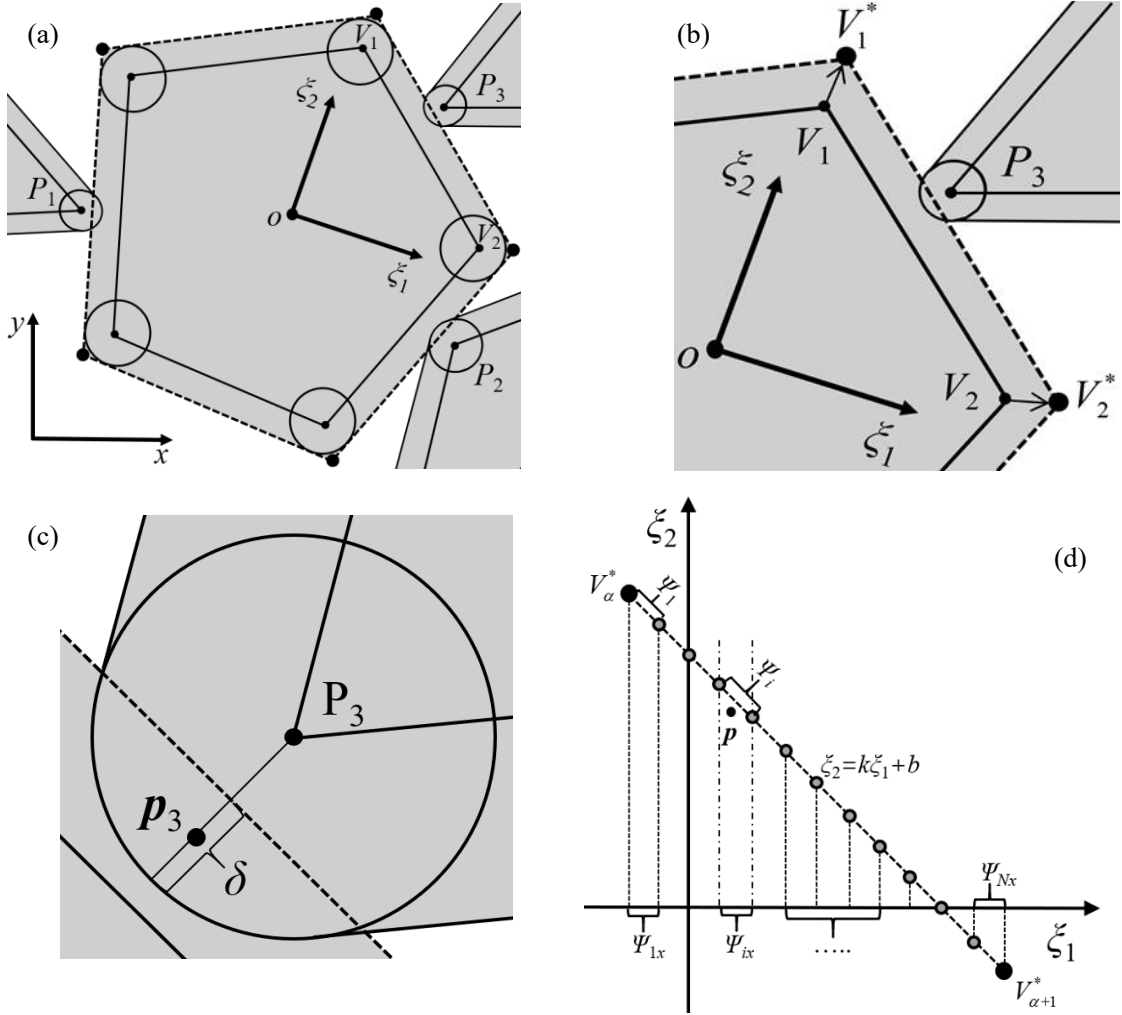

Figure 7 Procedures for importing contact forces into the traction vector; (a) particle with three vertex-edge contacts ($P_1$, $P_2$, $P_3$), where the solid and dashed line represent the original and dilated polygon, respectively; (b) a dilated edge defined by $V_1^*$, $V_2^*$; (c) the contact position $p_3$ of the contact $P_3$ calculated based on Eq. (20); (d) discretization of the particle edge ($V_\alpha^*, V_{\alpha+1}^*$). A contact position $p$ at the edge is linked with the corresponding boundary element $\Psi_i$ using its projection.

Figure 7 shows the boundary discretization and procedures to import the contact force into the boundary traction vector. The SDEM particle is first dilated. Points of intersection of dilated edges are used as new vertices. This dilation transfers the original spheropolygon to a standard polygon. Linear boundary elements are then generated on each edge defined by its dilated vertices (Figure 7(b)). The first and last elements coincide with the two vertices; therefore, no corner element needs to be used. In Figure 7(d), we use more general symbols to denote the contact position and its master edge. The contact position $p = (p_x, p_y)$ of a vertex-edge contact, which is calculated based on Eq. (20), and its force $f = (f_x, f_y)$ under the global coordinate system $(x, y)$ is first transformed to the particle's local coordinate system $(\xi_1, \xi_2)$ using the orientation angle of the particle. Then, the edge $(V_\alpha^*, V_{\alpha+1}^*)$ bearing the contact is decided by the minimum point-to-line distance (e.g., in Figure 7(b) and (c), contact $P_3$ will eventually be assigned to its closest edge defined by dilated vertices $V_1^*$ and $V_2^*$ for $\alpha=1$). Once the edge is decided, $p$ is assigned to a boundary element $\Psi_i$ based on its projection on $\Psi_{ix}$ at the local coordinate $\xi_1$ ($k<=1.0$) (or $\xi_2$ for $k>1.0$ with $\Psi_{iy}$). $p$ is always assigned to a smaller $\Psi_{ix}$ for $i>1$ if it coincides with the node of an element. Accordingly, the contact forces $f$ at $p$ are imported into the boundary traction component $t_i = (t_x = f_x/l_i, t_y = f_y/l_i)$ at element $\Psi_i$.

Each boundary element has two indexes: $I_{line}$ and $I_{global}$. The component $I_{line}$ represents the relative position of an element on the edge of the polygon and is used to link the contact position with the specific element on the closest edge. The index $I_{global}$ represents the global ID of an element in the traction vector $T$ in Eq. (12). Eventually, the contact $P$ is introduced into the traction vector $T$.

In this manner, the discontinuous method (SDEM) and the continuous method (BEM) are combined through the contact information. The position, velocity and contact forces of particles are simulated by SDEM, while the sub-particle stress is calculated by BEM using contact forces as boundary traction. The information of stress states could be further used to decide the damaged area and the shape of new particles for the replacement.

2.3.3 Criteria for the evaluation of sub-particle stress

Special conditions need to be fulfilled since SDEM is a dynamic method, while the BEM formulas we used for sub-particle stress calculation are only valid for the quasi-static state. This indicates that a particle should be in an equilibrium of force; its velocity for displacement and rotation also needs to be approximately zero before its sub-particle calculation is conducted. Therefore, the loading forces and velocities need to be examined before the calculation.

The first two inequalities, Eqs. (29) and (30), are used to guarantee a static state of the particle required by the BEM stress analysis:

$$[F_b = \sqrt{(\sum_{i=1}^n f_x^i)^2 + (\sum_{i=1}^n f_y^i)^2} \leq e_f] \wedge (|\tau_b| \leq e_\tau), \qquad (29)$$

where $r_{min}$ is the minimum radius of spheropolygon elements used in a simulation. The absolute values of the resultant force $F_b$ and torque $\tau_b$ are required to be smaller than the numerical tolerances $e_f$ and $e_\tau$. Ideally, $e_f$ and $e_\tau$ are equal to zero. In the programme, $e_f = 1 \times 10^{-6} r_{min} \times k_n$ and $e_\tau = (1 \times 10^{-3} r_{min})^2 \times k_t$ are scaled with contact properties ($k_n$, $k_t$), and the minimum radius to tolerate the computational errors (e.g., $1 \times 10^{-6}$ for single-precision variables). This condition guarantees a balanced force and torque on an individual particle. The sub-particle stress obtained without fulfilling this condition is invalid since the BEM formula used is only valid for calculating a quasi-static situation.

The second quasi-static condition is imposed on the kinematics:

$$(\sqrt{v_x^2 + v_y^2} \leq e_v) \wedge (v_a \leq e_{va}). \tag{30}$$

This inequality states that both velocity ($v_x$ and $v_y$) and angular velocity ($v_a$) should be smaller than $e_v = 1\times10^{-4}$ m/s and $e_{va} = 1\times10^{-4}$ rad/s. The values are also used as the tolerances of numerical errors. Because regularization is conducted, the results of sub-particle stress should not be affected by the absence of this condition. However, this condition is proposed with a practical meaning that a particle in high speed of motion or rotation is not going to break in a granular system because the BEM formula used here does not deal with a dynamic situation.

In a quasi-static granular system, only particles that are heavily loaded can potentially break and therefore need sub-particle stress calculation. Meanwhile, the repetitive BEM calculation should be avoided if the loading conditions on a particle do not change over time. The third and fourth condition, Eqs. (31) and (32), are introduced to examine a particle's loading magnitudes and loading history to only improve the computational efficiency. They do not affect the results of the simulation.

The third condition states that the contact forces must be sufficiently large, as otherwise the sub-particle stress will not lead to failure. This condition could be masked for one-way coupling BSEM.

$$(\sum_{i=1}^{n} |f_x^i| \geq f_{min}) \vee (\sum_{i=1}^{n} |f_y^i| \geq f_{min}). \tag{31}$$

This condition states that the sum of the absolute value of forces of each component ($f_x$ or $f_y$) needs to be larger than $f_{min}=0.1\times(A/\pi)^{1/2}\times K$, where $K$ is the tensile stress of the particle and $A$ is the area of the particle.

The fourth condition is used to avoid the repetitive calculation of a particle that does not experience enough change of contact forces ($\Delta F_x$ and $\Delta F_y$) after a certain number of time steps.

$$(\Delta F_x \geq \Delta F_{min}) \vee (\Delta F_y \geq \Delta F_{min}), \quad \Delta F_{min}=0.1\times F_{load}, \quad F_{load}=\frac{1}{2}(\sum_{i=1}^{n}|f_x^i|+\sum_{i=1}^{n}|f_y^i|). \tag{32}$$

The value of $F_{load}$ is updated at every step where all four conditions are fulfilled, and the calculation of sub-particle stress is conducted; its initial value equals $f_{min}$.

3. Single-particle test results

The BEM and BSEM share the exact same procedure for the single-particle stress calculation, where boundary forces are applied directly to the traction vector (Eq. (12)) instead of using inter-particle contact forces. Therefore, the accuracy and the efficiency of BEM, as the method for stress computation of BSEM, are tested in this part. The classic 2D Brazilian test is first presented, and the result of BEM is compared against the analytical solution. Then, the computational efficiency is compared with FEM results from ABAQUS. Next, a uniaxial compression test is executed with both circular and elliptic particles, and the full stress fields are obtained to demonstrate the ability of BSEM to calculate the stress state of irregular particles. Both force and torque equilibrium are fulfilled in each test. Only maximum tensile stress is measured and discussed. The results illustrate how abundant and insightful the sub-particle stress information that the well-established BEM can provide is.

3.1 Validation

A Brazilian test is chosen as a benchmark test. It is a widely recognized experimental method to measure the tensile stress of a rock specimen [32]. The classical Brazilian test involves uniaxial loads over a cylindrical sample. The analytical solution of the Brazilian test is well known. Its stress field shows that the maximum tensile stress occurs at the centre of the sample where the crack is initiated as a tensile failure. In particular, existing literature [4,7,10,33] suggests that the characteristic tensile stress induced by a pair of diametrical loads in a spherical grain is consistent with the maximum tensile stress of the Brazilian test. DEM breakage simulations also use the averaged stress tensor to approximate this result of the Brazilian test as a validation for the opposite-loading case [14].

The particle radius $R$=0.1 m is used in the calculations below. The material properties are as follows: shear modulus $G$=2.0×10$^3$ MPa and Poisson's ratio $v$=0.2. The same parameters are applied in all simulations.

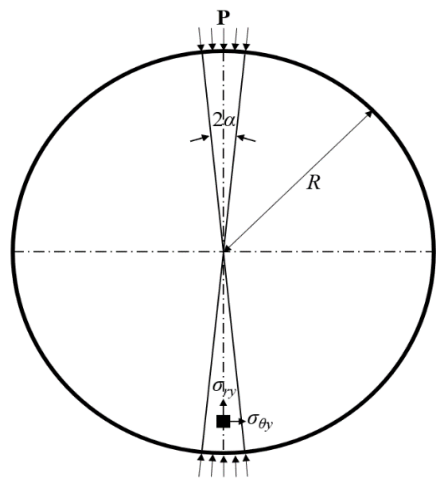

Figure 8 Brazilian disc test under a uniformly distributed load over finite arcs [32].

The standard Brazilian test is shown in Figure 8. The compressive pressure **P** is applied at the top and bottom of the sample. To prevent the sliding of the sample, the loading condition is usually not a pair of perfect point loads but rather a small contact area defined by an angle $\alpha$.

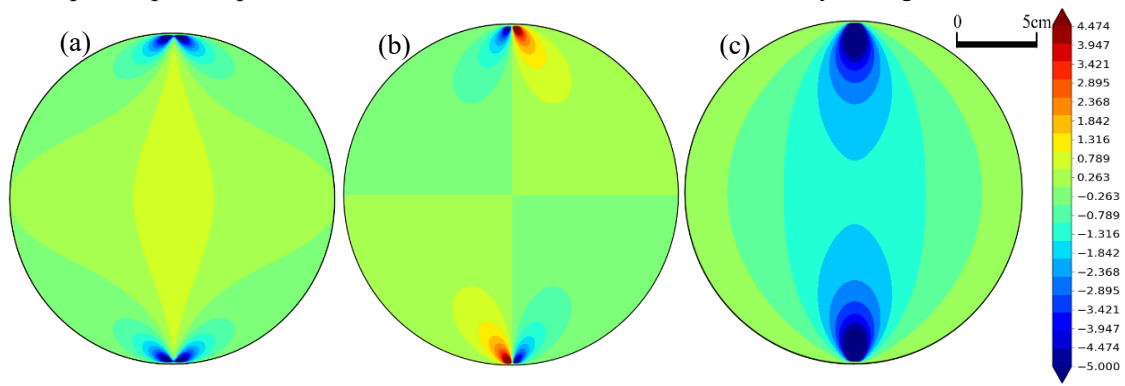

Figure 9 Stress components of the Brazilian test (Units: MPa); (a) $\sigma_{xx}$, (b) $\tau_{xy}$, (c) $\sigma_{yy}$.

The results of the BEM calculation are presented in Figure 9. BEM uses 223 linear boundary elements with a contact angle $\alpha=7.5°$. The resolution of inner stress points is 200 evenly distributed in the circumscribed square (or rectangle) of the particles. As we can see, the horizontal and vertical zero lines of shear stress are crossing the centre and are perpendicular to each other. The tensile zone in the middle of the disc and the stress field near the contact areas are also clearly presented. The BEM stress field shows good agreement with the analytical solutions. The tangential normal stress $\sigma_{\theta y}$ and radial normal stress $\sigma_{ry}$, as a function of the axial coordinate $r_y$, are given by [32]:

$$\sigma_{\theta y} = \frac{2P}{\pi} \left\{ \frac{(1-r_y^2/R^2)\sin 2\alpha}{1-2r_y^2/R^2 \cos 2\alpha + r_y^4/R^4} - \tan^{-1}\left[\left(\frac{1+r_y^2/R^2}{1-r_y^2/R^2}\right)\tan\alpha\right] \right\}, \tag{33}$$

$$\sigma_{ry} = -\frac{2P}{\pi} \left\{ \frac{(1-r_y^2/R^2)\sin 2\alpha}{1-2r_y^2/R^2 \cos 2\alpha + r_y^4/R^4} + \tan^{-1}\left[\left(\frac{1+r_y^2/R^2}{1-r_y^2/R^2}\right)\tan\alpha\right] \right\}, \tag{34}$$

The maximum tensile stress is obtained at the centre of the sample.

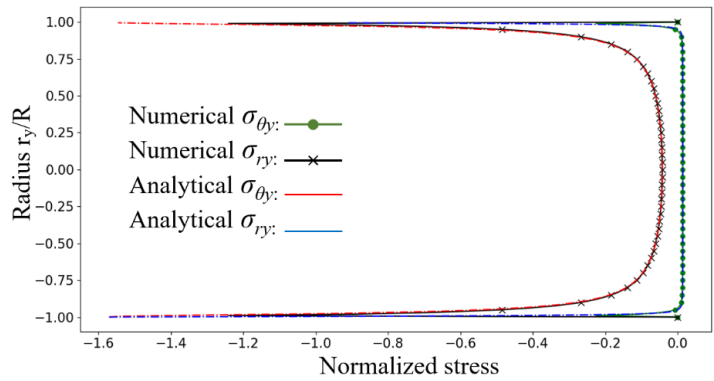

Figure 10 Axial stress distribution in the Brazilian test. The solid line denotes the analytical solution, and the dashed lines represent results from BEM.

A comparison of axial stress components is shown in Figure 10. The numerical results provided by BEM agree well with the analytical solution near the central area, while numerical errors can be observed near the boundary area as a result of the near-boundary singularity.

3.2 Performance analysis

Accuracy and efficiency comparisons between BEM and FEM using a second-order element in ABAQUS [15] are shown in Figure 11. The error is calculated by comparing the tensile stress at the centre of the disc with that of the analytical solution.

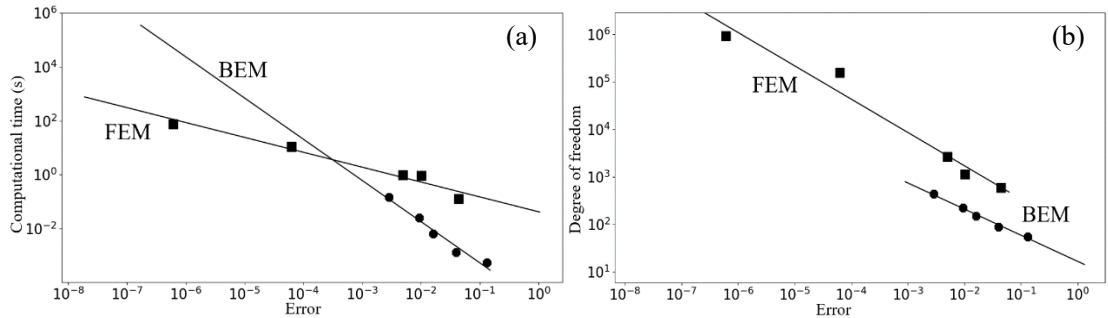

Figure 11 Computational performance of BEM compared to FEM (ABAQUS); (a) computational time versus numerical error; (b) degree of freedom versus numerical error. The solid lines are generated with first-order polynomial regression. BEM uses constant elements, whereas FEM uses quadratic elements.

The performance of both cases is obtained from the Brazilian test with an increasing number of mesh (boundary and body) elements. It can be observed from Figure 11(a) that, for the same accuracy, BSEM has better efficiency than FEM. The computational time is directly linked to the size of the coefficient matrix and the number of degrees of freedom (DOF). For an equal consumption of computational time, BSEM is more accurate. In contrast, BSEM is less accurate for errors below $2\times10^{-4}$ that correspond to the value where the two lines in Figure 11(a) cross. This indicates that BEM is a more suitable method for a low number of DOF. To achieve equal accuracy, a smaller size of the matrix (i.e., lower number of DOF) is more favourable. In this case, particles do not need to be finely meshed, while a continuous and accurate stress field can still be obtained. This advantage of BEM is more significant if the simulation involves a larger number of particles.

It must be pointed out that the relations between CPU and degrees of freedom are algorithm- and processor-dependent. However, the exponents of the power law are inherent for both methods. The computational times compared here are only the 'solver times,' i.e., the CPU times for solving the matrix. It excludes (or at least minimizes) the differences of implementations, such as the time a programme spends on discretization, IO flows and other pre-processing procedures before solving the matrix. Therefore, the relative relationship of efficiency, especially under low DOF, is still valid.

3.3 Sub-particle stress calculation

A series of parametric studies of sub-particle stress are presented in this section. The effects of two variables for the breakage, aspect ratio ($AR$) and coordination number ($CN$), are investigated with BEM. The sub-particle stress fields are calculated and analysed. The heterogeneity of the sub-particle stress for both circular (the same model used for the Brazilian test) and elongated elliptical grains (the number of linear boundary elements ranged between 230 to 250) are compared. Then, the validity of three formulas for particle strength is examined: the effect of aspect ratio, cushioning effect, and averaged stress tensor. The purpose of this section is to provide a new perspective on the influence of particle shape and meanwhile demonstrate the limitation of the stress averaging method, which are widely accepted and adopted in numerous studies of particle fragmentation.

3.3.1 Aspect ratio

After particle size, the second most important geometrical variable that affects the breakage of a particle is aspect ratio (*AR*). The characteristic tensile stress $\sigma$ for a circular grain is a formula of diametric force *F* and its radius *d*, which can be written as [4,33]:

$$\sigma = \frac{F}{d^2}. \tag{35}$$

As mentioned earlier, this formula is consistent with the analytical solution of maximum tensile stress in a Brazilian test and can be used for the measure of the tensile strength of grains. If the particle is non-circular, Eq. (35) needs to be modified. Based on experimental data, Afshar et al. [31] proposed the following equation:

$$\sigma = \frac{F \times AR}{D^2}, \tag{36}$$

where *AR=D/d* is the aspect ratio; *D* denotes the sieve-measured diameter of the particle, and *d* is the laminar thickness. This formula suggests that, with an equal pair of loads, particles with larger *AR* would have larger tensile stress. Thus, the tensile strength of elongated particles measured or observed would appear to be lower. This modification of the strength formula was proposed by empirical research without providing a theoretical explanation.

With the aim of testing Eq. (36), we conducted BEM tests on a single particle using spheres and ellipses with different aspect ratios. *AR* is defined here as the ratio of the length of the major and minor axes (b/a). *AR*=1.0 is chosen for circular particles with radius *r*=0.1 m, and *AR*>1.0 is used for all of the elliptical particles. These geometrical parameters are applied to all of the relevant tests. *d* is the distance between two loading points; the areas of the particles are set to be equal to eliminate the statistical effect of the Weibull distribution [33]. Particles are equally loaded with a pair of axial compression pressures, which resembles the Brazilian test. The maximum tensile stress is measured for each case. The diametrical load is equal to $2\times10^5$ N.

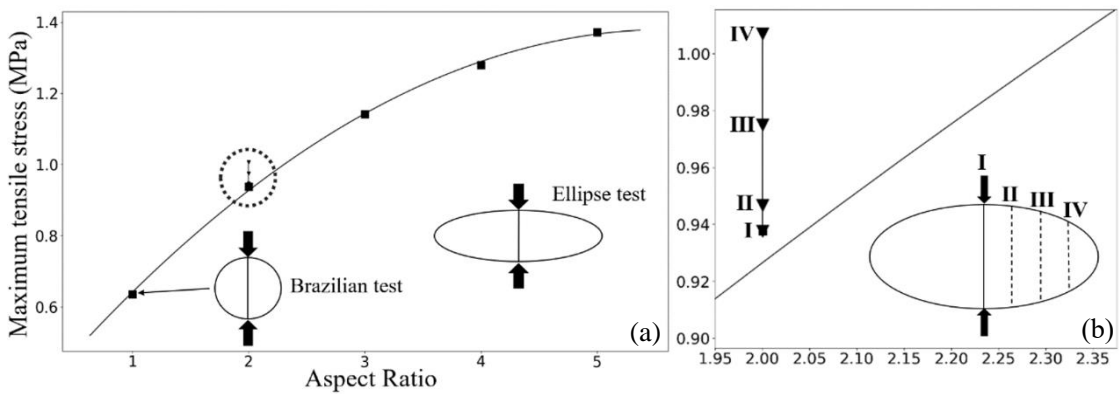

Figure 12 (a) The maximum variation of the stress with aspect ratio using second-order polynomial regression; *AR*=1.0 is a 2D Brazilian test; *AR*>1.0 for tests with elliptical particles; (b) amplified circular area of *AR*=2.0 with different loading positions.

The curve in Figure 12 (a) shows that the maximum stress in a particle increases with *AR*. The elliptical particles with larger *AR* are easier to break than circular particles. It is not reasonable that the macroscopic strength can be changed by the shape. As the results suggest, the so-called *AR*-strength effect is caused by the maximum tensile stress, which is directly related to the shape of the particle. Eq. (36) is thus a linear approximation of the dependency of tensile stress with the aspect ratio of the particle, but non-linearities should be expected.

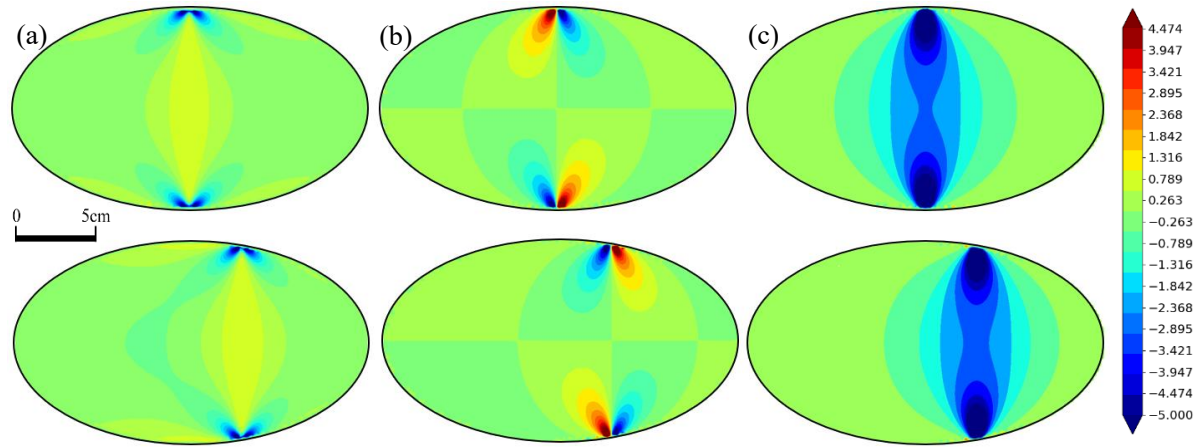

Figure 13 Stress fields of an elliptical particle with *AR*=2.0 and the loading positions I and II (Units: MPa); (a) $\sigma_{xx}$, (b) $\tau_{xy}$, (c) $\sigma_{yy}$.

The second test is conducted to further understand this effect: the same loading condition is applied at different positions of an elliptical particle with *AR*=2.0, and the loading positions move from the centre towards the tip of the particle with abscissa *x*=0, *b*/4, *b*/2, 3*b*/4. Figure 12 (b) indicates that the maximum tensile stress increases with the decrease of the distance between the two contact points. It can be well explained with fundamental solutions where the intensity of the stress field is strongly affected by the distance-dependent terms $\ln(1/r)$, $1/r$ or $1/r^2$. For a circular particle, two stress fields overlap with each other and produce the maximum tensile stress at the exact centre. Such overlapping becomes stronger if the distance decreases on particles that have larger *AR* values, or one particle but at different positions. Although the physical meaning is not accurate, the modification in Eq. (36) is consistent with fundamental solutions. It is therefore reasonable to use laminar thickness to modify the formula for the strength of a particle with *AR*>1.0.

3.3.2 Coordination number

Particle breakage is related to the loading condition, which includes the following three key variables: the load intensity, the central angle of applied loads, and the contact number. The latter is also called the coordination number (*CN*). McDowell et al. [4] assumed that increasing *CN* further decreases the chance of breakage. Although they acknowledged that such an assumption is insufficient or even invalid for elongated particles, it has been embedded into many DEM or analytical models of particle breakage. It is reasonable to assume that a large value of *CN* (e.g., *CN*>10) may prevent the breakage of a particle due to the cushioning effect [4] (many small particles are in contact with one big particle such as a cushion). However, for a low number of contacts, such an assumption is questionable and requires further examination.

A series of BEM tests are conducted to investigate the effect of low *CN* on the sub-particle stress. As shown in Figure 14, circular (*AR*=1.0) and elliptical particles (*AR*=2.0) are used for the comparison between different *AR*s. Four groups of contact angles are applied with *CN*=3, 4.

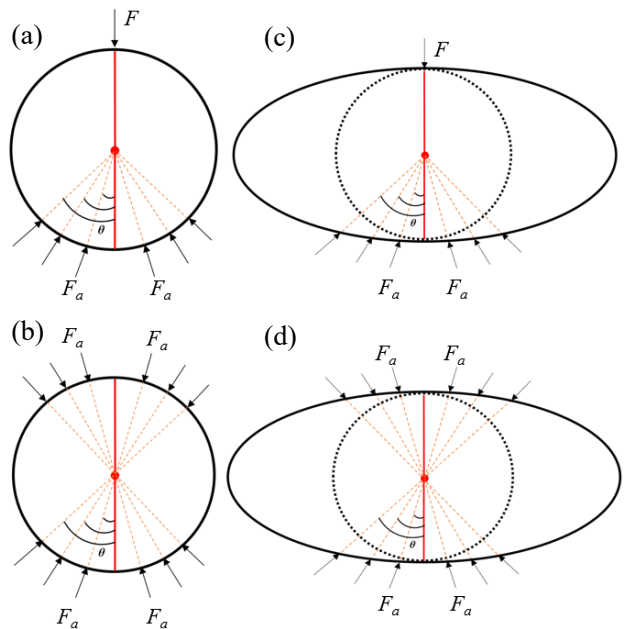

Figure 14 The loading conditions of circular and elliptical particles, where the imposed loads are: $F=2\times10^5$ N, $F_x=\pm 0.5F\tan(\theta)$, $F_y=-0.5F$, $F_a=(F_x^2+F_y^2)^{1/2}$, $\theta = 0°, 15°, 30°, 45°$. (a) and (b) *CN*=3; (c) and (d): *CN*=4.

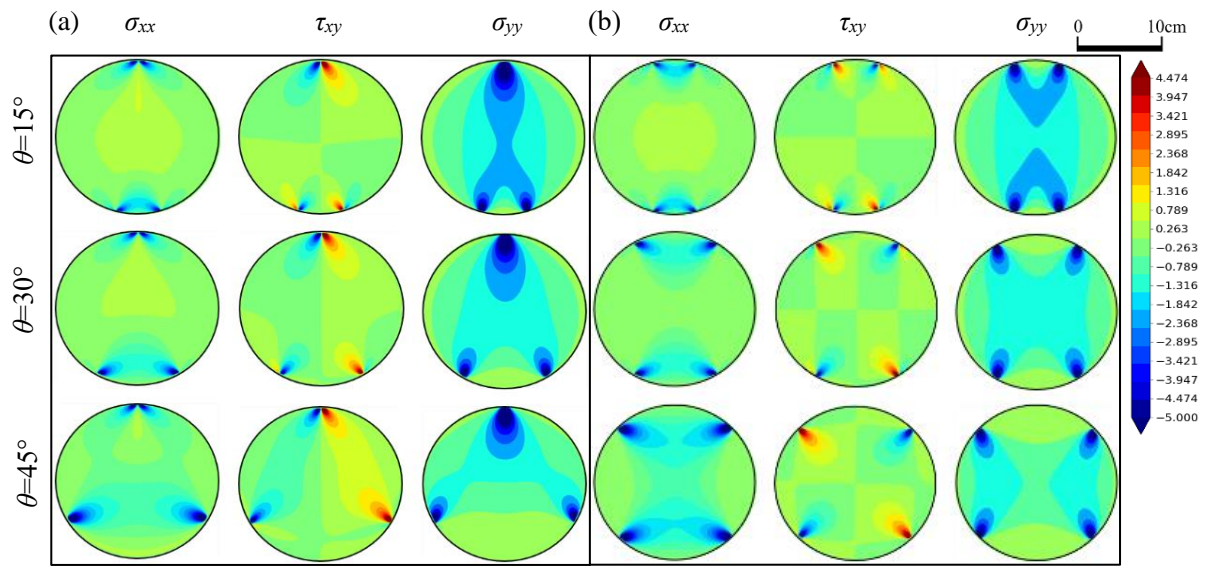

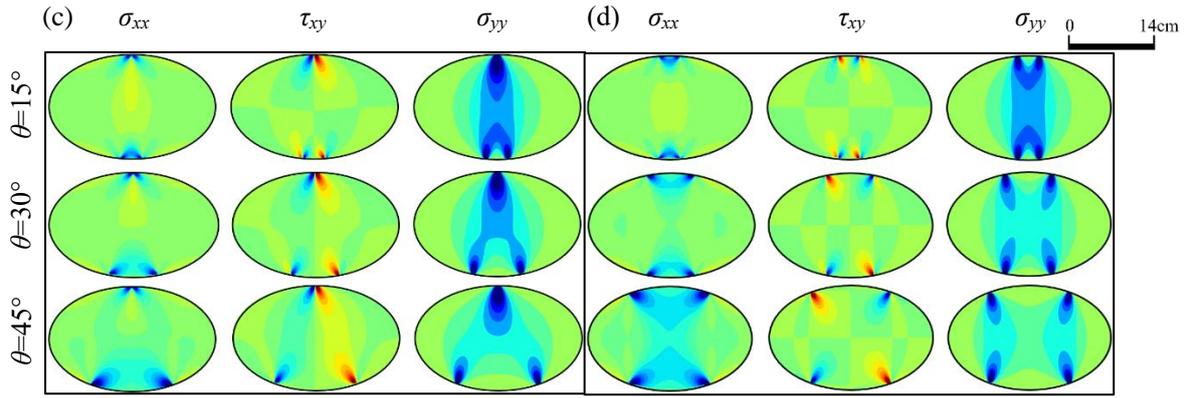

Figure 15 Stress components for each loading case (a)-(d) in Figure 14 (Units: MPa). The results of different loading angles are arranged as $\theta$=15°, 30°, 45° from top to bottom and stress $\sigma_{xx}$, $\tau_{xy}$, $\sigma_{yy}$ from left to right.

The stress fields in Figure 15 show that the sub-particle stress has a strong dependency on the loading angle. For the circular particle, the maximum overlapping area shrinks with increasing loading angles, which also causes changes of tensile failure position and maximum value.

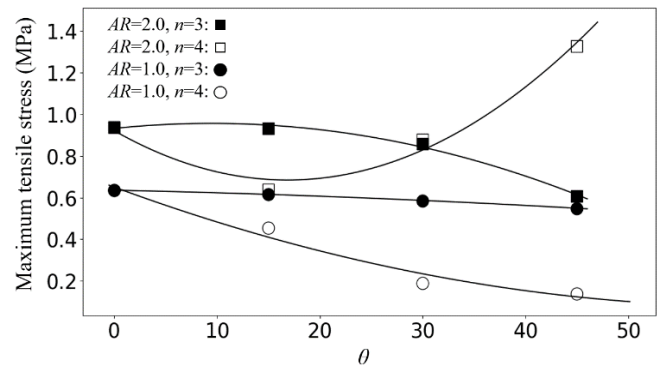

Figure 16 The maximum tensile stress with different loading angles and coordination numbers.

It can be observed from Figure 16 that the effect of the coordination number agrees well with the existing assumption regarding the cushioning effect for circular particles. $CN$=4 has a lower maximum tensile stress than that of $CN$=3; the drop of maximum tensile stress is also consistent with it. However, the above is not valid for the elliptical particles, where the shape of the particle distorts stress fields. The maximum tensile stress even increases with contact angle or the value of $CN$. Such inconsistency is directly caused by the shape of the particle and usually fails to be properly considered by many studies.

3.3.3 Heterogeneity of sub-particle stress

The average stress tensor is the foundation for many of the numerical and analytical methods. It can represent the stress state in a particle without consideration of its irregular geometry. As we discussed in the earlier parts through the analysis of maximum tensile stress, this tensor shows good agreement with the Brazilian test in circular particles, but it is unable to accurately capture the stress state in the elliptical particle. The heterogeneity of the sub-particle stress is further investigated here using circular and elliptical models ($AR$=2.0) subject to an equal pair of loads.

The maximum tensile stress along the different diametrical directions is measured.

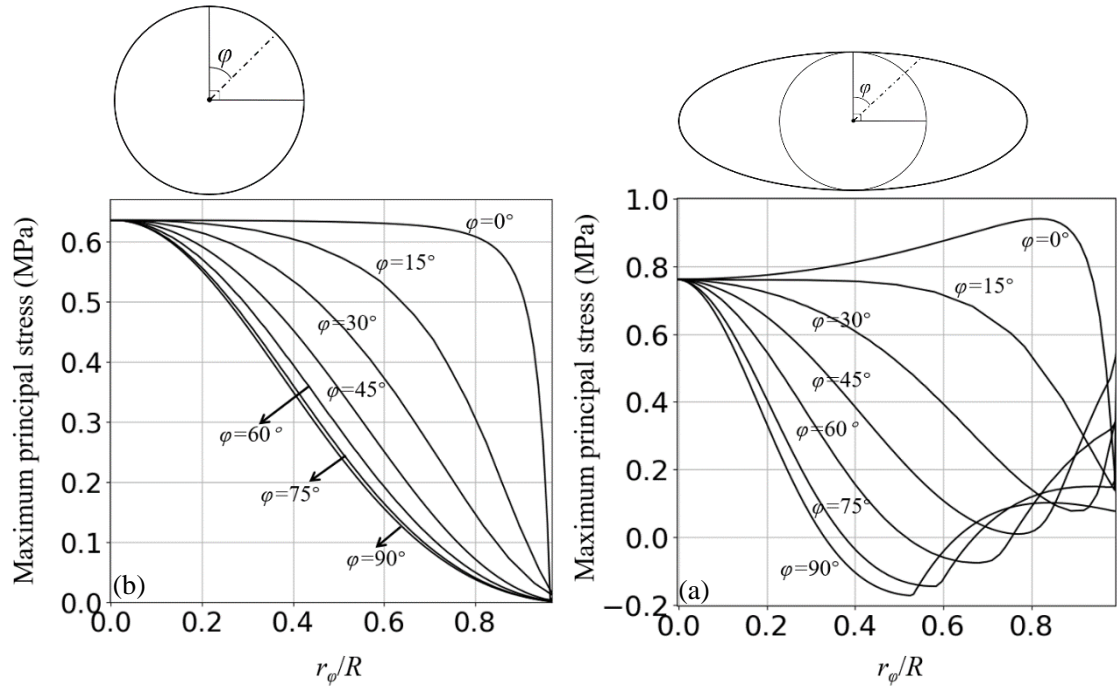

Figure 17 Maximum tensile stress distribution at diametrically opposed directions with different angles; (a) circular particle case, (b) elliptical particle ($AR =2.0$) case.

The maximum tensile stress is plotted diametrically ($r_\varphi$ is the distance to the centre, and $R$ is the distance between the centre and the boundary) at the circular and elliptical particles with different angles. Figure 17 shows that, for the elliptical particle, the heterogeneity of the stress is stronger than that of the circular particle. The maximum tensile stress can still be found at the central line but is no longer located at its centre point. It also shows a more drastic variation along some of the diametrical directions with a range that includes both negative and positive values. Therefore, the validity of an average tensor is highly compromised for the irregular particle simulation. Neither the breakage time nor the breakage shape can be properly predicted by using the averaging method.

3.3.4 Discussion

The analysis of the tensile stress in the previous section provides three important highlights regarding how the description of the particle breakage can be improved, as follows: (1) the aspect ratio is an essential particle shape parameter to include in the particle strength criteria; (2) the increase of the coordination number does not necessarily lead to a reduction of particle breakage as the cushioning effect suggests; and (3) the average stress tensor is not sufficient to predict breakage, and the stress heterogeneities need to be taken into account. The breakage of an individual particle is a complicated process that is affected by a variety of conditions, such as contact forces, material properties, and geometry. The sub-particle stress components govern the distribution of damaged areas and the development of breakage interfaces, and they cannot be reduced to a universal equation. It is not accurate enough to simply average the overall stress states of a particle and ignore the heterogeneity of the stress field. Therefore, the sub-particle stress provides a crucial perspective for a better understanding of particle breakage. It is desirable for a numerical simulation of particle breakage to be governed by accurate stress information.

4. **Granular model tests of BSEM**

In this section, we use a one-way BSEM coupling method to simulate the uniaxial compression test of granular materials. The tests are performed to demonstrate the ability of BSEM to provide insightful information on sub-particle stress distribution in a granular system. For all of the tests, particles are placed in a container that consists of three rigid walls at right, left and bottom. The compression force is applied by a rigid bar with a constant magnitude $F_y$=-2×10$^5$ N. The calculation of sub-particle stress is conducted after the force equilibrium (Eq. (29)) on each particle is achieved, and the entire granular system is in a quasi-static state (Eq. (30)). The specific number of boundary elements on each particle is proportional to its perimeter. This is due to the rule that the length of the element is equal to the radius of the spheropolygon, as discussed in Section 2.3.1. Using a sphero-radius of $r$=1.0×10$^{-3}$ m, this rule generates less than 300 elements on the particles simulated here. The gravity is excluded from both the calculation of sub-particle stress and the SDEM simulation. The material properties and parameters of the simulations are listed in Table 1.

Table 1 Parameters used in the simulation

| | | | |
|---|---|---|---|
| $k_n$ | Normal stiffness | 6.0×10$^8$ | N/m |
| $k_t$ | Tangential stiffness | 1.2×10$^7$ | N/m |
| $\mu$ | Friction coefficient | 0.5 | |
| $g$ | Gravitational constant | 0.0 | m/s$^2$ |
| $\rho$ | Density | 2.5 | g/cm$^3$ |
| $\Delta t$ | Time interval | 10$^{-5}$ | s |
| $V_d$ | Verlet distance | 0.5 | cm |
| $G$ | Shear modulus | 2×10$^3$ | MPa |
| $v$ | Poisson's ratio | 0.2 | |

Figure 18 shows how BSEM is conducted for circular and angular particles. It can be seen from the continuity of the stress fields that the contact information provided by SDEM is properly coupled with the traction boundary conditions in BEM. In particular, the stress fields near the contact point of two particles are symmetrical. This indicates that both the magnitude and the position of the contacts are correctly imported into the traction vector and therefore provide a reliable result for the stress field on each particle. Furthermore, the stress fields for angular particles show a stronger concentration and heterogeneity of the contact forces than those of circular particles.

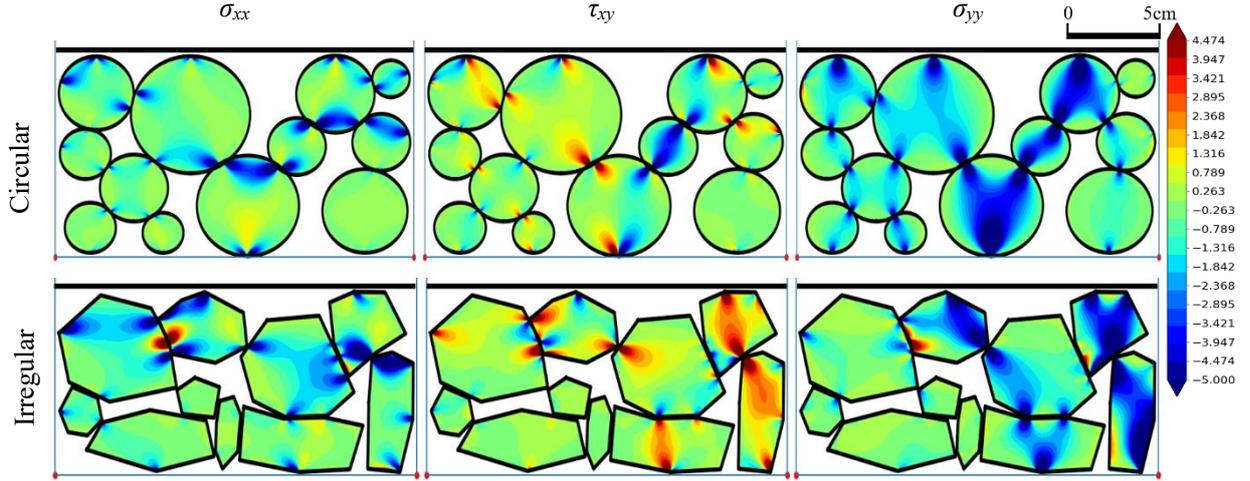

Figure 18 Sub-particle stress of the circular and irregular particles (Units: MPa).

With the aim of further investigating the effect of particle circularity, we perform different simulations by keeping the void ratio approximately constant and only changing the particle circularity. Simulations are performed with disks and regular polygons (dodecagons, decagons, and octagons). The following formula for the circularity is adopted:

$$CR=\frac{4\pi A}{B^2}, \qquad (37)$$

where $A$ is the area, which is controlled as a constant value ($A=4\pi$ for the circular particle's radius $R=2.0$ cm) for all of the particles; $B$ is the perimeter of the particle, which will increase with the decrease of circularity. $CR$ equals 1.0 for a circular particle and decreases with the increase of the number of angles. Only three layers of particles are used to fully exhibit the details of the sub-particle stress fields. A sphero-radius of $r=1.0\times10^{-3}$ m is also applied here and produces less than 200 elements for each particle. Particles are arranged with random orientations at the beginning of each simulation to further minimize the influence of the packing pattern.

It can be observed from the sub-particle stress fields in Figure 19 that the loading forces for circular particles are symmetrical. This is because the compressive load from the bar is evenly distributed to the particles. Particularly, the stress component $\sigma_{yy}$ has the highest magnitude for each particle as a reaction to the vertical compression. The stress distribution is consistent with the results provided by Zheng and Tannant [12], who used force chains of agglomerates to represent the sub-particle stress. The results are also consistent with the loading test on single circular particles in Section 4.2, and it further confirms that the sub-particle stress fields in the granular material cannot be represented by the average stress tensor, especially when particles have multiple contacts.

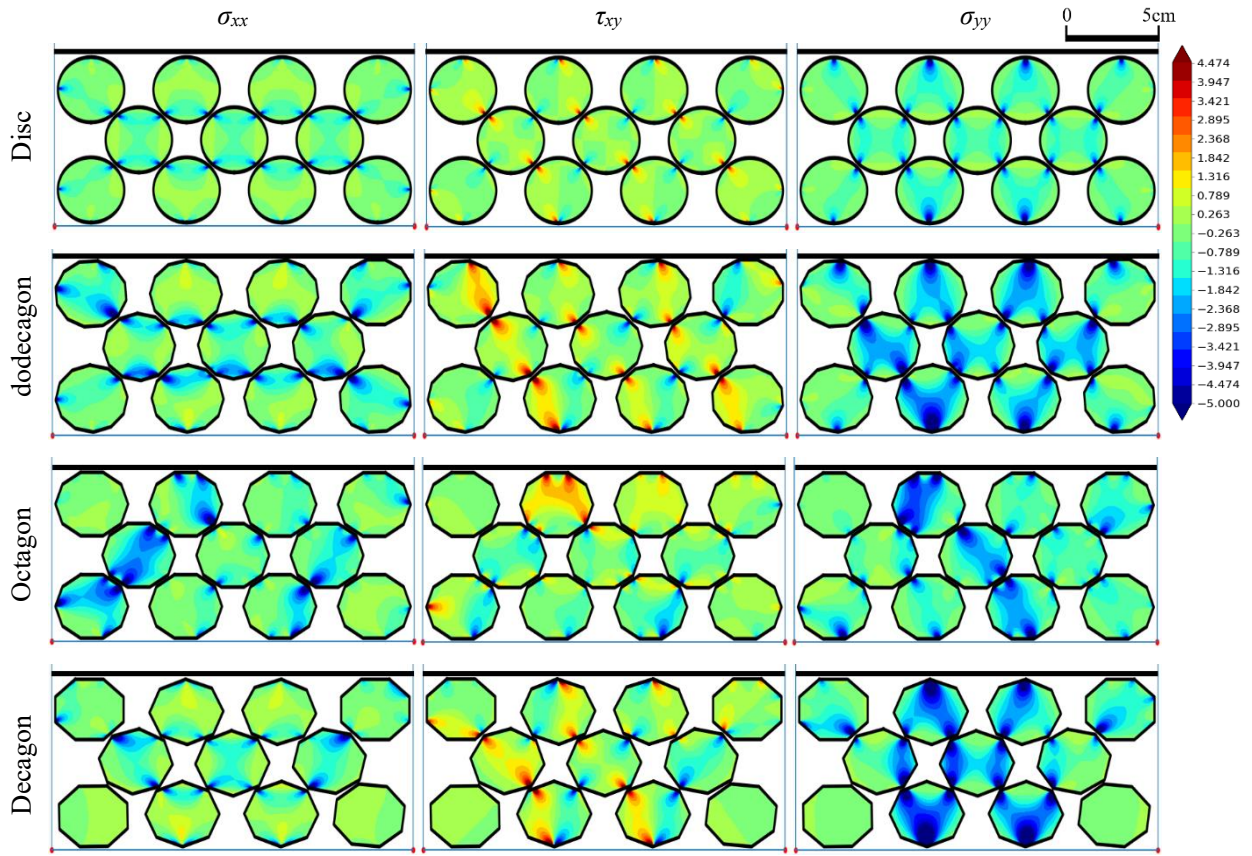

Figure 19 Sub-particle stress components in discs and angular (dodecagon, decagon, and octagon) particles (Units: MPa).

The patterns of sub-particles stress fields for angular particles in Figure 19 show that the decrease of the circularity enhances the heterogeneity of the contact network. Additionally, such a decrease causes the loading conditions on particles to become less even and no longer symmetrical. Some of the particles bear most of the loads, while the others are much less or not loaded. The stress component $\sigma_{yy}$ still has the highest value as a reaction to the vertical compression, yet the magnitudes of $\sigma_{xx}$ and $\tau_{xy}$ are increased due to the uneven distribution of forces and changes of particle geometry. The numerical comparison between the results of different circularity (*CR*) models also confirms that the heterogeneity of the force distribution increases with the decrease of circularity.

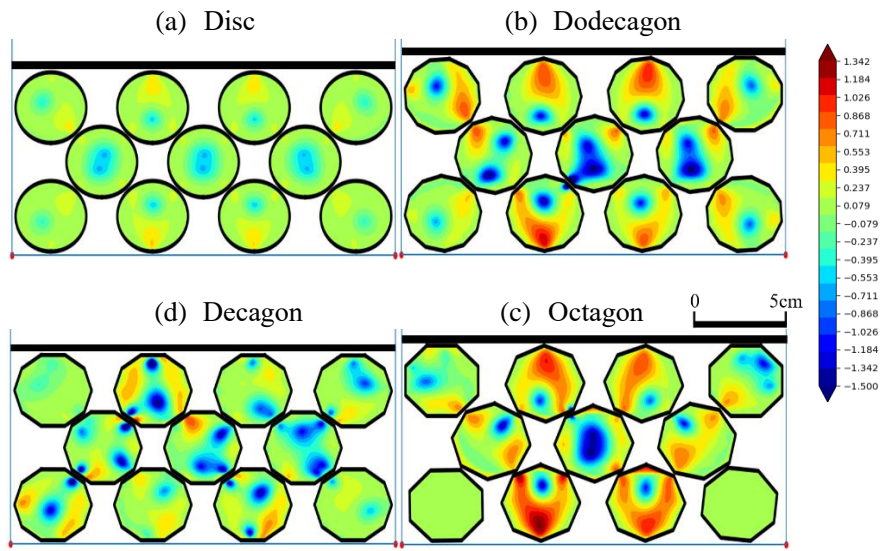

Figure 20 Sub-particle maximum principal stress of the particle models with different circularity (Units: MPa).

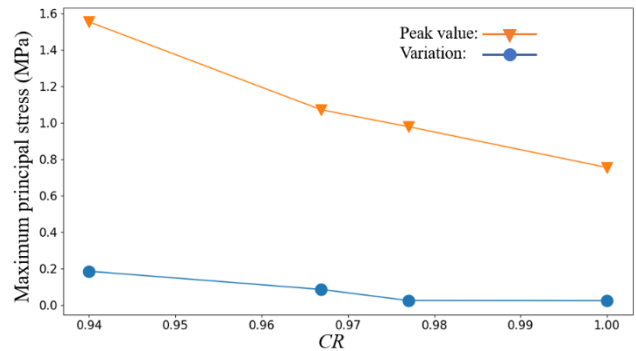

Figure 21 The peak value (PV) and averaged variation of the maximum principal stress (AVM) as a function of circularity.

The values of the maximum principal stress field for the particles in each model are shown in Figure 20; the peak value (PV) and averaged variation of the maximum principal stress (AVM) of the entire granular packing are shown in Figure 21. The stress field in the system of the circular particle is symmetrical, and it also has the lowest variation and peak value. Both PV and AVM are largely increased with the decrease of $CR$, which is consistent with the stress fields in Figure 19. Generally, the circularity of the particles directly affects the distribution of the force chain. The heterogeneity of the force chain makes the particles unevenly loaded and produces complicated sub-particle stress fields. In such a situation, more breakage will be observed in the model with lower circularity. It must be pointed out that the circularity of the particle does not directly decide the breakage resistance for an individual particle but instead changes the heterogeneity of the force chain in a granular assemble and produces more breakage.

## 5. Conclusions

An advanced numerical method, BSEM, has been developed for the simulation of the particle stress state, which will be used to model particle breakage in the future. It combines the advantages of the boundary element method and the spheropolygon element method. BSEM can effectively capture both the geometrical irregularity of particles and their interaction forces. The sub-particle stress can be accurately calculated using only meshes at the particle's boundary. Performance analysis using the Brazilian test indicates that the stress fields provided by BSEM are more accurate and computationally less expensive than FEM-related methods for low degrees of freedom. A domain mesh is not required, and thus breakage interfaces and the damage zone are not limited or affected by pre-defined elements. It has been pointed out that the computational time of the BSEM is considerably increased if a high-resolution stress field is calculated. However, a full stress field does not need to be calculated to simulate particle breakage. The stress state only needs to be considered at certain areas or lines on a particle, while the continuity and accuracy can still be maintained by using fundamental solutions in a better fashion than the shape function of FEM, which produces stress discontinuities. In the former case, the computational time for a high-resolution stress field can be largely reduced or even neglected.

Parametric studies using BSEM demonstrate advantages over existing formulas that are widely used in numerical methods by showing the discrepancy between the actual stress state and the results produced by the oversimplified assumptions that are generally applied in simulations. An advanced understanding of particle breakage requires accurate sub-particle stress fields. Further simulations of the uniaxial loading test demonstrate that BSEM can provide insightful information on the sub-particle stress distribution. It indicates that decreases in the circularity of particles increase the heterogeneity of the force chain and hence cause high tensile stress areas in certain particles. This effect will lead to more breakage in the granular material and shows a lower breakage resistance for more angular particles.

It can be concluded that BSEM overcomes most of the existing problems in numerical methods for evaluation of the stress states of particles. It provides a superior balance of accuracy and efficiency. BSEM has a great potential to instruct stress-based methods to simulate breakage interfaces and further improve the replacement strategy for broken particles.

While our method is still a one-way coupling method that does not include particle breakage, the analysis of the sub-particle stress provides important insights on the roles of particle aspect ratio and coordination number in particle breakage. The full-coupling method requires the incorporation of failure criteria and energetic principles to replace the broken particle, and this will be discussed in future publications. We anticipate that BSEM will make a good contribution to the study of particle breakage in granular materials.

Acknowledgments

We thank Falk Wittel for critical discussions and for producing the ABAQUS calculations in Figure 11 of this paper. We also thank the reviewers for their valuable comments, all of which helped us to improve the quality of this paper.